\documentclass{appolb}
\usepackage{graphicx}

\begin{document}

\title{Prospects for Pentaquark Baryon Search with the Upgraded LEPS2 Facility%
}

\author{
T.~Nakano, S.~Ajimura, Y.~Asano, S.~Dat\'e, T.~Hashimoto, A.~Higashi, T.~Hotta, T.~Ishikawa, H.~Katsuragawa, R.~Kobayakawa,
H.~Kohri, K.~Mizutani, Y.~Ohashi, H.~Ohkuma\thanks{deceased},
S.Y.~Ryu, S.~Suzuki, S.~Tanaka, K.~Watanabe, B.~Yan,
T.~Yorita, M.~Yosoi
\address{Research Center for Nuclear Physics, Osaka University, Ibaraki, Osaka 567-0047, Japan}
\\[3mm]
G.~Kojima, M.~Miyabe, N.~Muramatsu, H.~Ohnishi, Y.~Sada,
H.~Shimizu, A.O.~Tokiyasu
\address{Research Center for Accelerator and Radisotope Science, Tohoku University, Sendai, Miyagi 982-0826, Japan}
\\[3mm]
M.~Niiyama, K.~Nishi
\address{Department of Physics, Kyoto Sangyo University, 
Kyoto 603-8555, Japan}
\\[3mm]
J.K.~Ahn
\address{Department of Physics, Korea University, Seoul 02841, Republic of Korea}
\\[3mm]
Y.~Ma
\address{
RIKEN Cluster for Pioneering Research, RIKEN, Wako, Saitama 351-0198, Japan}
\\[3mm]
T.H.~Nam
\address{Dalat Nuclear Research Institute, Dalat, Lam Dong 66106, Vietnam}
\\[3mm]
C.~Rangacharyulu
\address{Department of Physics and Engineering Physics, University of Saskatchewan, Saskatoon, Saskatchewan S7N 5E2, Canada}
\\[3mm]
M.~Sumihama
\address{Department of Education, Gifu University, Gifu 501-1193, Japan}
C.~Yoshida
\address{
Institute of Particle and Nuclear Studies, High Energy Accelerator Research Organization, Tsukuba, Ibaraki 305-0801, Japan}
}

\maketitle

\begin{abstract}
We present prospects for the $\Theta^+$  pentaquark baryon search using the newly constructed LEPS2 facility at SPring-8. The LEPS2 detector system features significant improvements in acceptance for multi-particle final states compared to previous experiments. Our search employs two complementary strategies: direct  production  in the $\gamma n \to K^-\Theta^+$ reaction using a liquid deuterium target with a photon beam up to 2.4 GeV, and $\bar{K}^{*0}$-associated $\Theta^+$ production using a liquid hydrogen target with a photon beam up to 2.9 GeV. The extended acceptance covers both forward and large angle regions, effectively spanning the kinematic regions explored by previous LEPS and CLAS experiments. The large acceptance and improved resolution of LEPS2, combined with these complementary approaches, provide unprecedented sensitivity for establishing the existence of the $\Theta^+$ or placing definitive upper limits on its production.
\end{abstract}

\section{Introduction}

The existence of exotic hadrons beyond the conventional quark model has been a fundamental question in hadron physics since Gell-Mann's prediction of multiquark states \cite{Gell-Mann:1964ewy}. Among these exotic states, the $\Theta^+$ pentaquark baryon ($uudd\bar{s}$) has attracted particular interest due to its unusually narrow width and light mass, which challenge our understanding of QCD dynamics.

The $\pi$, $K$, and $\eta$ mesons, as Nambu-Goldstone bosons corresponding to the spontaneous breaking of chiral symmetry in QCD, generally facilitate rapid decay of multi-quark states. This mechanism typically leads to large widths for exotic hadrons. However, if relatively stable multi-quark states exist, they would provide crucial insights into fundamental degrees of freedom beyond the constituent quarks in hadron descriptions.

The renewed interest in pentaquark baryon searches was initiated by the theoretical work of Diakonov, Petrov, and Polyakov \cite{Diakonov:1997mm}, who predicted the $\Theta^+$ as a member of the anti-decuplet with mass around 1.53 GeV and width less than 15 MeV. The experimental journey began with the first observation at LEPS in 2003 \cite{LEPS:2003wug}, where the $\Theta^+$ was observed in the $\gamma n \to K^-\Theta^+$ reaction with a mass of $1540 \pm 10$ MeV/$c^2$. This discovery was initially supported by observations at several facilities, including the DIANA collaboration's analysis of $K^+$Xe collisions \cite{DIANA:2003uet}, which suggested an exceptionally narrow width of $\Gamma(\Theta^+) = 0.34 \pm 0.10$ MeV/$c^2$.

The experimental landscape has evolved considerably since these first observations. High-statistics experiments at the CLAS facility in JLab \cite{CLAS:2006czw} failed to confirm their earlier positive results. The J-PARC E19 experiment \cite{J-PARCE19:2014zgo} set stringent upper limits of 0.28 $\mu$b/sr in the $\pi^-p \to K^-X$ reaction at 1.92 GeV/$c$. These null results have led to intense scrutiny of the experimental methods and theoretical understanding of pentaquark production mechanisms.

Theoretical analyses have led to refined predictions for the $\Theta^+$ properties. The chiral soliton model predicts a mass of 1530 MeV/$c^2$, while QCD sum rules suggest a mass in the range of 1500--1600 MeV/$c^2$ \cite{Nishikawa:2004su}, \cite{Gubler:2009iq}. Lattice QCD calculations typically predict higher masses above 1600 MeV/$c^2$. The width predictions vary wildly, from as narrow as 0.36 MeV/$c^2$ to as broad as 15 MeV/$c^2$. A comprehensive review of these theoretical predictions can be found in \cite{Praszalowicz:2024zsy} and \cite{Amaryan:2022iij}.

The discovery landscape for exotic hadrons has been transformed by the observation of hidden-charm pentaquarks at LHCb \cite{LHCb:2019kea}. The observation of $P_c(4312)^+$, $P_c(4440)^+$, and $P_c(4457)^+$, with widths ranging from 6.4 to 20.6 MeV, has shown that pentaquark configurations can indeed be realized in nature. However, these states have different characteristics from the proposed $\Theta^+$, especially with respect to their decay widths and production mechanisms.

\section{LEPS2 Experimental Facility}

The LEPS2 facility represents a significant upgrade from the original LEPS experiment, specifically designed to address the challenges in the spectroscopy of exotic hadrons. The facility uses two different experimental setups with different targets and beam energies: one with a liquid deuterium target using a photon beam up to 2.4 GeV, and the other with a liquid hydrogen target using a higher energy photon beam up to 2.9 GeV.

A major technical advance in LEPS2 is its high-intensity photon beam production system \cite{LEPS2beamline}. Thanks to simultaneous injection of laser light from multiple sources and the synchronization between the pulsed laser photons and electron bunches in the storage ring, LEPS2 achieves tagging intensities up to 5 MHz, with typical experimental operation in the range of 1.5--1.8 MHz. This represents a significant improvement in beam intensity over the previous facility, enabling the accumulation of highly statistical data sets necessary for the study of rare processes.

The detector system has been optimized to achieve high performance in several critical aspects. The extended acceptance covers both forward and wide angle regions, providing nearly complete coverage of the reaction phase space. A key feature of LEPS2 is its large acceptance for multi-particle final states, which is particularly important for reactions with three or more particles in the final state.

The momentum resolution for charged particles varies with the detector region: about 2\% in the forward region using TPC and DC tracking, and about 4\% in the backward region using TPC tracking alone, at 1 GeV/c. The facility also features efficient neutral particle detection, which complements the charged particle measurements and allows complete event reconstruction. The mass resolution of about 6 MeV/$c^2$ for the $K^0p$ system represents a significant improvement over previous experiments, crucial for the observation of potentially narrow states.

The high-intensity polarized photon beam at LEPS2 provides additional tools to study the production mechanisms and quantum numbers of all observed states. The beam intensity and quality, combined with the detector capabilities, allow the accumulation of highly statistical data sets necessary for detailed analysis of rare processes.

\section{Search Strategy}

\begin{figure}[htbp]
    \centering
    \includegraphics[width=8cm]{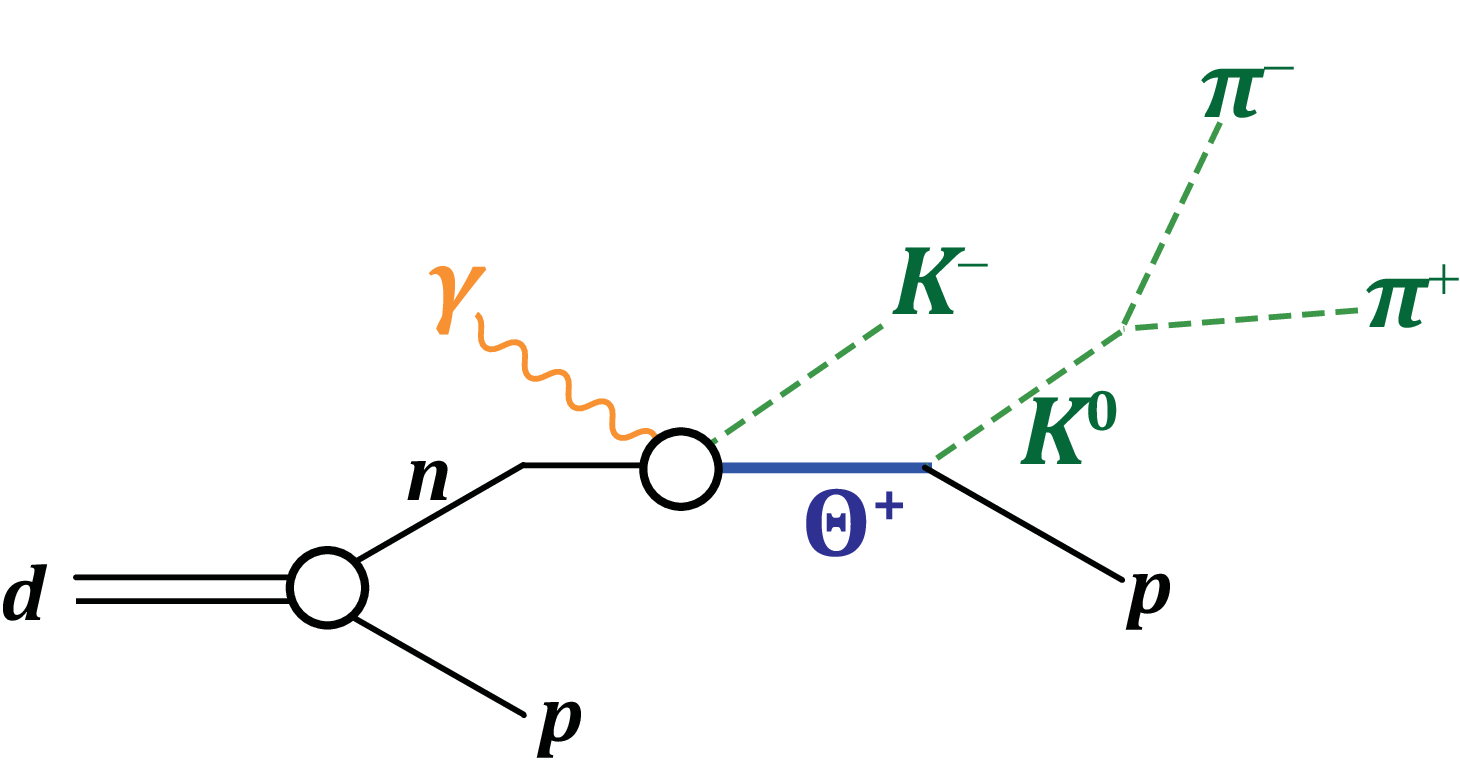}
    \caption{Conceptual diagram of the  $\gamma n \to K^-\Theta^+$ reaction followed by the $\Theta^+ \to pK^0$ decay}
    \label{fig:fig1}
\end{figure}

Our search for the $\Theta^+$ employs two complementary approaches. The first approach uses the deuterium target to study the $\gamma n \to K^-\Theta^+$ reaction, where $\Theta^+$ can be observed through both the $pK^0$ and $nK^+$ decay modes. The $pK^0$ channel offers particular advantages through the clear identification of the $K^0_S \to \pi^+\pi^-$ decay, providing strong kinematic constraints for background suppression.  Figure \ref{fig:fig1} shows a conceptual diagram of the $\gamma n \to K^-\Theta^+$ reaction. Complete reconstruction of all final state particles allows precise determination of the $\Theta^+$ mass and width.

\begin{figure}[htbp]
    \centering
    \includegraphics[width=8cm]{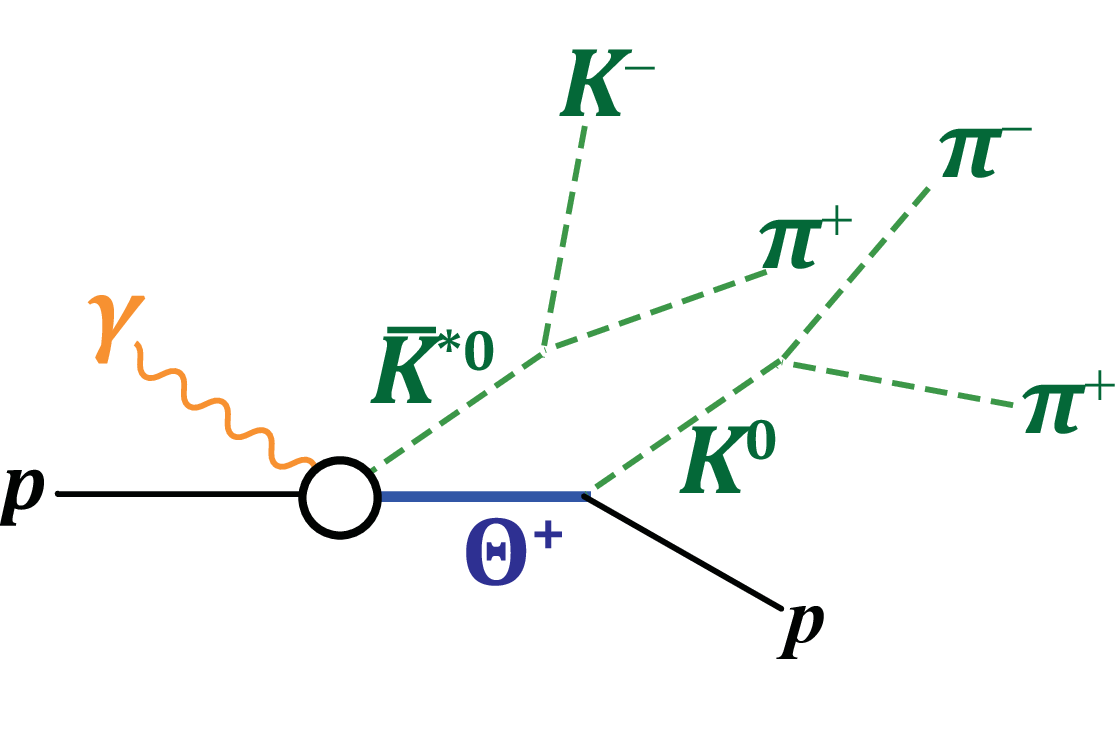}
    \caption{Conceptual diagram of the  $\gamma p \to \bar{K}^{*0}\Theta^+$ reaction}
    \label{fig:fig2}
\end{figure}

The second approach, using the hydrogen target at higher beam energies, explores the $\bar{K}^{*0}$-associated $\Theta^+$ production via the  $\gamma p \to \bar{K}^{*0}\Theta^+$ reaction. This channel offers several significant experimental advantages due to its unique characteristics. The use of a hydrogen target eliminates complications from Fermi motion effects, resulting in improved mass resolution for the $\Theta^+$ reconstruction. The reaction topology is particularly clean due to the well-defined $\bar{K}^{*0} \to K^-\pi^+$ decay, which provides strong kinematic constraints through the known $\bar{K}^{*0}$ mass. These features, combined with the ability to study angular distributions in detail, provide enhanced sensitivity to the underlying production mechanism.  Figure \ref{fig:fig2} shows a conceptual diagram of the $\gamma p \to \bar{K}^{*0}\Theta^+$ reaction. 

Background suppression is achieved through multiple complementary techniques specific to each channel. The $pK^0$ decay mode offers a significant advantage over previous LEPS measurements investigating the $nK^+$ channel, as it is free of background contributions from $\phi$ meson production. This is particularly important because $\phi$ meson production was a major source of background in the original LEPS experiment, complicating the analysis of the $nK^+$ channel. In the hydrogen target measurements with $\bar{K}^{*0}$ production, the $\bar{K}^{*0}$ mass constraint provides strong background rejection. 

The polarized photon beam allows detailed studies of spin observables, which are crucial for determining the quantum numbers of each observed state. Measurements of the beam asymmetry will constrain the spin-parity assignment of the $\Theta^+$. Furthermore, potential interference between $\Theta^+$ production and the $\phi$ meson production processes will provide an additional tool for studying the production mechanism and coupling strengths.

\section{Expected Performance}

\begin{figure}[htbp]
    \centering
    \includegraphics[width=8cm]{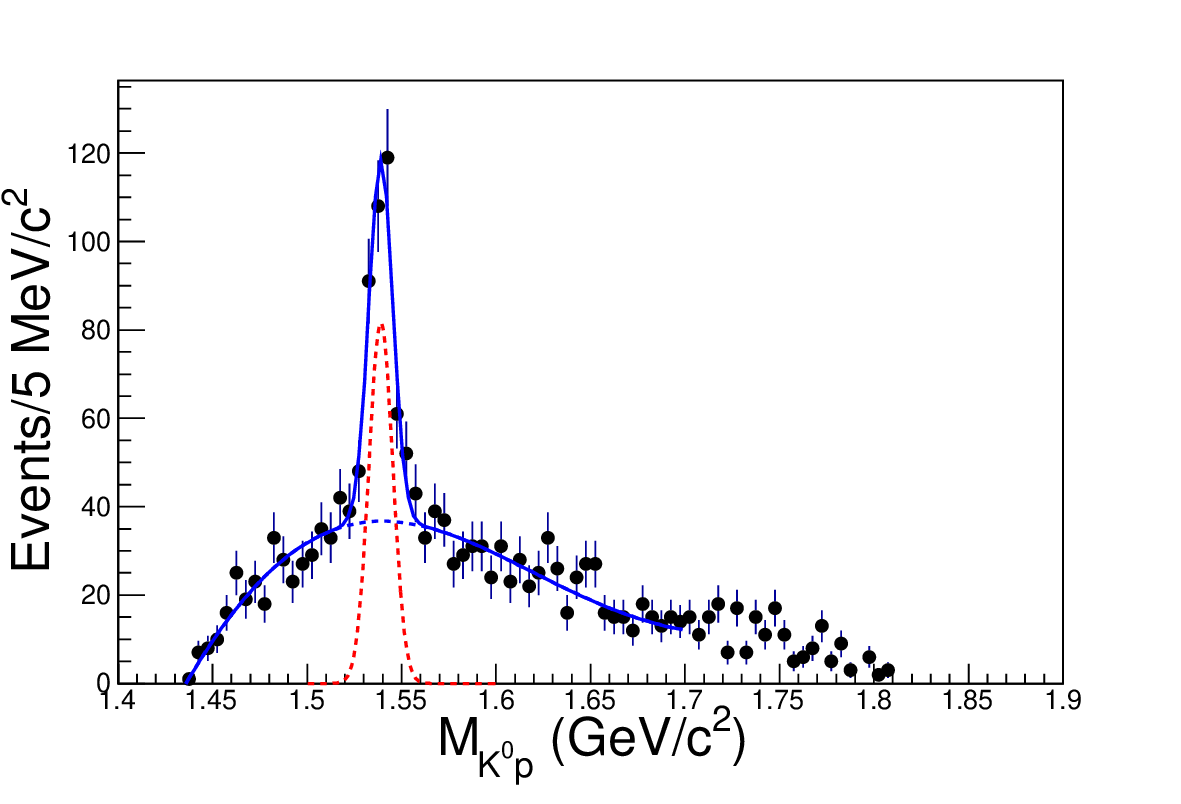}
    \caption{$pK^0$ invariant mass distribution obtained through the simulation study, assuming a $\Theta^+$ production cross section of 4 nb/str.}
    \label{fig:fig3}
\end{figure}

The performance expectations for the $\Theta^+$ search at LEPS2 are based on detailed Monte Carlo simulations that include all major reaction channels and detector effects. These simulations include the full detector response, reconstruction algorithms, and analysis procedures that will be used in the actual experiment. For the deuterium target measurements, with a data acquisition rate of 100 cps, we expect to accumulate about $8.6 \times 10^8$ events during 100 days of beam time. Figure \ref{fig:fig3} shows the $pK^0$ invariant mass distribution obtained from the simulation study, assuming a $\Theta^+$ production cross section of 4 nb/str.

Simulation studies are ongoing for the hydrogen target measurements exploring the $\bar{K}^{*0}\Theta^+$ channel. The geometric acceptance for the $\gamma n \to K^-\Theta^+$ reaction, considering events where $K^-$ is emitted within 20 degrees in the forward direction, is about 10\%. This acceptance is mainly determined by the detector coverage and the reaction kinematics.

Background studies using simulations have revealed several critical aspects of the measurement. The primary background sources include non-resonant $KN$ production, $\phi$ meson production with associated strange particles, and combinatorial backgrounds from multiparticle final states. Each of these backgrounds exhibits distinct kinematic features that can be exploited for their suppression. The relative contributions of these backgrounds differ significantly between the two measurement approaches, allowing for complementary background studies and cross-checks of the background subtraction procedures.

The high statistics and comprehensive angular coverage of the experiment will allow sophisticated analysis techniques beyond simple mass spectrum studies. Detailed partial wave analysis will separate potential signals from background contributions, using both the angular distributions of the decay products and the beam polarization observables. This analysis will be particularly powerful in determining the spin and parity of each observed state, since different $J^P$ assignments lead to different predictions for the angular distributions and polarization observables.

\section{Conclusions}

The LEPS2 facility represents a significant advance in the experimental capabilities for exotic hadron spectroscopy. Through the combination of high beam intensity, excellent detector resolution, and nearly complete angular coverage, this experiment is uniquely positioned to resolve the long-standing question of the existence of the $\Theta^+$ pentaquark baryon. The complementary approaches using both deuterium and hydrogen targets provide independent paths to search for this state, each with its own advantages and systematic considerations. The sensitivity to cross sections at the nanobarn level, combined with excellent mass resolution and sophisticated analysis capabilities, ensures that LEPS2 will either provide a definitive observation of the $\Theta^+$ or set stringent limits that will significantly impact our understanding of exotic hadron states in QCD.

\section*{Acknowlegments}
The authors thank the following colleagues
for contribution of fruitful discussions on physics promotion and
construction of the LEPS2 beamline,
and R\&D works of detectors and circuits: 
A.~Averyanov,
W.C.~Chang,
J.Y.~Chen,
M.L.~Chu,
Y.~Furuta,
H.~Goto,
T.~Hiraiwa,
K.~Horita,
H.~Ikuno,
Y.~Kasamatsu,
H.~Kawai,
D.~Krivenkov,
M.~Kubo,
K.~Miki,
R.~Mizuta,
Y.~Narita,
K.~Nakamura,
T.~Nakamura,
T.~Nobata,
K.Y.~Noh,
T.~Noritake,
Y.~Nozawa,
T.~Ohta,
R.~Richa
H.~Saito,
T.~Shibata,
S.~Shibuya,
R.~Shirai,
E.A.~Strokovsky,
M.~Tabata,
Triloki,
M.~Tsuruta,
E.~Umezaki,
Y.~Yanai,
R.~Yamamoto.
and
H.M.~Yang.
The use of the BL31LEP of SPring-8 (the LEPS2 beamline) has been approved by the Japan Synchrotron Radiation Institute (JASRI) as a contract beamline (Proposal Nos. BL31LEP/6101 and 6102).
This work has been supported in part by JSPS KAKENHI Grant Numbers JP15H00839, JP16K13806, JP17H02892, JP20H01933, 21H01110, 21H01115,  JP21H04986, JP22H00124, JP23H01201, and 24H00232.


\end{document}